\documentclass[fleqn,twoside]{article}
\usepackage{espcrc2}
\usepackage{graphicx}
\newcommand{\AmS}{{\protect\the\textfont2
  A\kern-.1667em\lower.5ex\hbox{M}\kern-.125emS}}

\title{A new method for the UHECR mass composition studies}
\author{M.Ambrosio\address{INFN, section of Napoli, Napoli, Italy}
       \thanks{E-mail: Michelangelo.Ambrosio@na.infn.it},
       C.Aramo$^{\rm a,}$\address{Dept. of Physical Sciences, Univ. of Napoli
 Federico II, Napoli}, 
       D.D'Urso\address{Dept. of Physics and Astronomy, Univ. of
 Catania and INFN, Section of Catania, Catania, Italy},  
       A.D.Erlykin$^{\rm a,}$\address{P.N.Lebedev Physical Institute, Moscow, 
Russia}, F.Guarino$^{\rm a,b}$, A.Insolia$^{\rm c}$}

\begin{document}

\begin{abstract}
\noindent The determination of the primary cosmic ray mass
composition from the longitudinal development of atmospheric
cascades is still an open problem. In this work we propose a new method of the  
multiparametric topological analysis and show that if both $X_{max}$ - the depth of 
shower maximum and $N_{max}$ - the number of charged particles in the shower maximum 
are used, reliable results can be obtained.
\end{abstract}

\maketitle

\section{Introduction}

The study of the longitudinal profile of individual
atmospheric cascades started in the early eighties with the
development of the fluorescent light detection technique
implemented for the first time within the framework of the Fly's
Eye experiment \cite{Baltr}. After these
pioneering efforts, there has been only one experimental array
continuing this type of studies: HiRes \cite{Matth} and
only recently a new and much more powerful detector has started to
collect data: the Fluorescent Detector (FD) of the Pierre Auger
Observatory \cite{Bluem}. This instrument will produce a
large data flow over the next decades and is therefore calling for
new and accurate data analysis procedures capable to fully exploit 
the large amount of information contained in the FD data.

It is rather surprising, that while there are many methods, both parametric
and non-parametric (KNN, Bayesian methods, pattern recognition,
neural nets etc.), used to discriminate individual cascades on the
basis of ground--based information, very little has been done to fully
exploit the amount of information contained in FD data. To
our knowledge, the only method available so far makes use of $X_{max}$
 - the depth of the maximum cascade development \cite{Gaiss}
and derives the observed {\em mean} mass
composition as a function of the primary energy. Since there is a
minimum bias in the detection of cascades of different origin, the
observed mass composition coincides practically with the primary
composition. It has to be stressed also that this approach relies
on statistical grounds and therefore does not allow the
identification of the primary particle for each individual
cascade. Furthermore, even though in the longitudinal profile the
$X_{max}$ parameter is the most sensitive to the mass of the
primary particle, its sensitivity is still weak. For instance, at
a primary energy of 1 EeV (10$^{18}$eV) the mean iron induced
cascade has $X_{max}$ only (11-12)\% lower than for a proton
induced one, i.e. a difference which is of the same order of
magnitude as the intrinsic fluctuations of $X_{max}$.

As we shall discuss below, such unsatisfactory situation
improves drastically if other, seemingly less significant
parameters, are taken into account. Among them, we might have:
$N_{max}$ - the number of particles (mostly electrons) in the
maximum of the cascade, the speed of rise in the particle number
etc. For instance, at fixed primary energy, $N_{max}$ is about the
same for all cascades, and even though iron induced cascades
produce more muons and less energy is carried out by the
electrons, the effects on $N_{max}$ are very small. However, due
to the lower energy per constituent nucleon in the primary iron nucleus,
the cascade development and the rise of the cascade curve are on
average faster than for cascades originating from protons. 
This useful information is neglected when only $N_{max}$ is taken into
account. This type of arguments triggered our efforts to find
methods capable to fully exploit the information contained in the
cascade curves and to allow the identification (at least in terms
of probabilities) of the cascade origin also for individual
showers.

Although tailored for possible
applications in the context of the Pierre Auger experiment,
 methods described
below are quite general and may find application in other similar experiments.
However, in what follows we shall focus on data similar to those
expected from the Fluorescent Detector (FD) of the Pierre Auger
Observatory: namely on the longitudinal profile of each atmospheric
cascade, i.e. on the number of charged particle $N_{ch}$ as a function of the
atmospheric depth $X$. As it will be shown below, this
profile carries more information than $X_{max}$ or $N_{max}$
alone.

We would like to stress that even though the fraction of the hybrid
events, in which the information on the shower from the Pierre Auger
Surface Array is supplemented by the Fluorescent Detector data,
will hardly exceed 10\% of the total statistics accumulated by the Surface Array, 
these events
need to be properly handled since they contain the maximum
information. In this paper we restrict ourselves to the
analysis of the longitudinal development of cascades.
The treatment of hybrid data will be addressed in subsequent papers.

\section{The simulated data}

In what follows, we assume that the primary energy
estimates for hybrid events will be accurate at a few percent level \cite{PAPDR}.
The data set used to implement and test the
methods described in the following sections consists of 7600
vertical cascades produced by particles with the fixed energy of 1
EeV, simulated using the CORSIKA program (version 6.015, \cite{Heck}) 
with the QGSJET interaction model. Simulations were performed at the Lyon Computer Centre. 
The primary nuclei were $P$, $He$, $O$ and $Fe$, each of them initiating 1900
cascades. The CORSIKA output provides the number of charged
particles at atmospheric depths sampled with 5 gcm$^{-2}$ intervals.

We clipped the data at a depth of 200 g
cm$^{-2}$, since the FD detection threshold does not allow to
detect the weak signals at the beginning of the cascade
development. The maximum atmospheric depth was set at 870
g cm$^{-2}$, roughly corresponding to the level
of the Pierre Auger Observatory. In Figure 1
we show a subsample of 50 cascades for each primary.
\begin{figure}[htb]
\begin{center}
\includegraphics[width=7.5cm,height=7.5cm,angle=0]{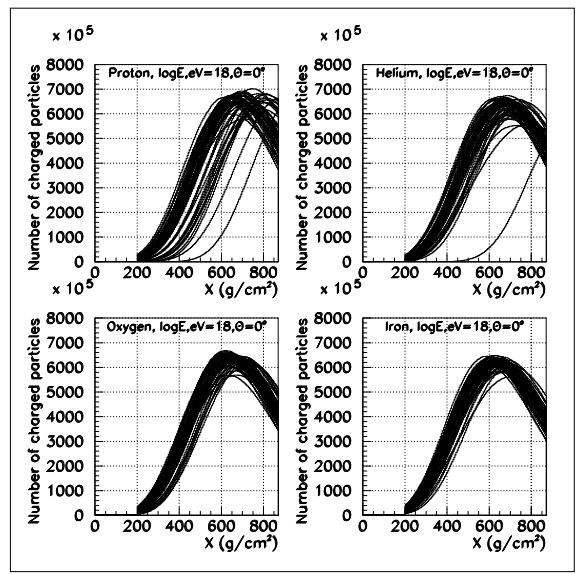}
\caption{\footnotesize A few examples of the longitudinal
development of 1 EeV vertical atmospheric cascades induced by
protons P, helium He, oxygen O and iron Fe nuclei, specified in headers. 
Each plot shows 50 typical cascades.}
\end{center}
\label{fig:mta1}
\end{figure}

\section{Multiparametric Topological Analysis (MTA) }

\begin{figure}[htb]
\begin{center}
\includegraphics[width=7.5cm,height=7.5cm]{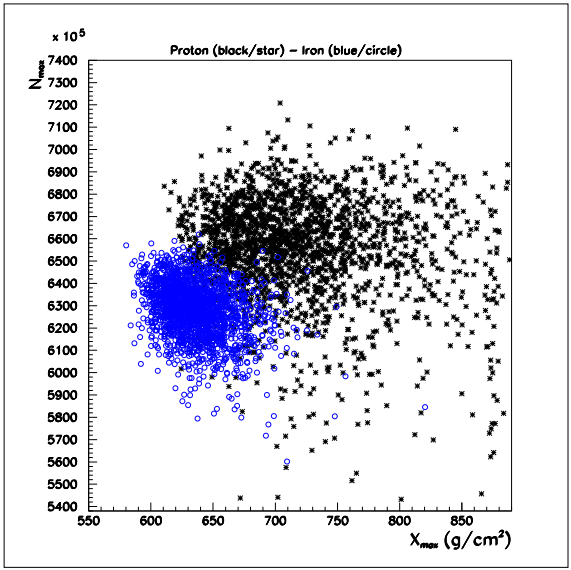}
\caption{\footnotesize N$_{max}$ vs. X$_{max}$ scatter plot for
proton (star) and iron (circle) induced showers}
\label{fig:mta2}
\end{center}
\end{figure}

\begin{figure}[htb]
\begin{center}
\includegraphics[width=7.5cm,height=7.5cm]{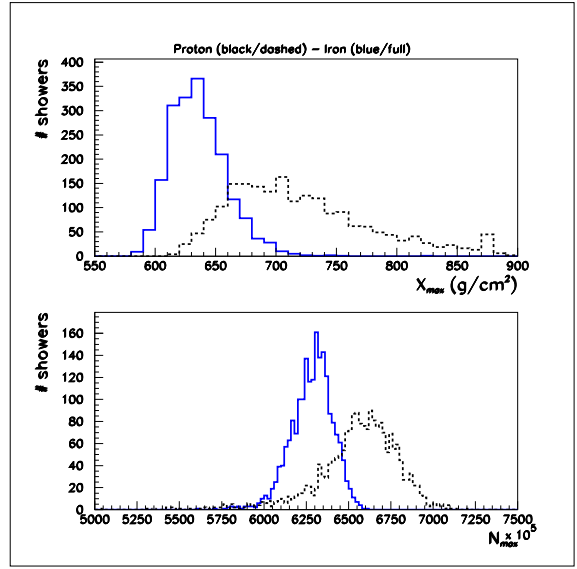}
\caption{\footnotesize Projected distributions of the scatter plot
in Figure \ref{fig:mta2} for protons (dashed line) and iron nuclei (full line)}
\label{fig:mta3}
\end{center}
\end{figure}

The MTA method, when applied to FD data, relies on a
topological analysis of correlations between the most significant
parameters of the shower development in the atmosphere.
In principle this method could be used also with a greater number of parameters,
however, in this paper we restrict ourself to the simple case of two parameters only: 
X$_{max}$ (the atmospheric depth of the shower maximum) and N$_{max}$ 
(the number of charged particles at the depth of
X$_{max}$). A scatter plot of these two parameters has been built using the
$4 \times 1000$ showers already described. Figure \ref{fig:mta2}
shows the scatter plot for only proton and iron induced showers
while their projected distributions are shown in Figure \ref{fig:mta3}.

The basic idea of the MTA method is to divide the
entire area of the scatter plot into cells whose dimensions are defined by an
accuracy with which the parameters can be measured. In our simulations
the value of 20 g/cm$^2$ was used as the width of X$_{max}$
bin, while 50$\times 10^5$ was assumed for the width of N$_{max}$ bin. 
In each cell we can find the total number
of showers N$_{tot}^{i}$, as the sum of N$_P^{i}$, N$_{He}^{i}$,
N$_O^{i}$ and N$_{Fe}^{i}$ showers induced by P, He, O and Fe
respectively, and then derive the associated frequencies:
p$_P^{i}$=N$_P^{i}$/N$_{tot}^{i}$, p$_{He}^{i}$=N$_{He}^{i}$/N$_{tot}^{i}$, 
p$_O^{i}$=N$_O^{i}$/N$_{tot}^{i}$ and
p$_{Fe}^{i}$=N$_{Fe}^{i}$/N$_{tot}^{i}$ which can be interpreted
as the probability for a real shower falling into the $i^{th}$ cell to
be initiated by proton, helium, oxygen or iron primary nuclei.
In other words, in the case of an experimental data set of $N_{exp}$
showers, it may be seen as composed of a mixture of N$_{exp}$ x
p$_P$ proton showers, N$_{exp}$ x p$_{He}$ helium showers,
N$_{exp}$ x p$_O$ oxygen showers and N$_{exp}$ x p$_{Fe}$ iron
induced showers, where p$_A = \Sigma_i$p$_A^i$.

\begin{figure}[h]
\begin{center}
\includegraphics[width=7.5cm,height=7.5cm,angle=0]{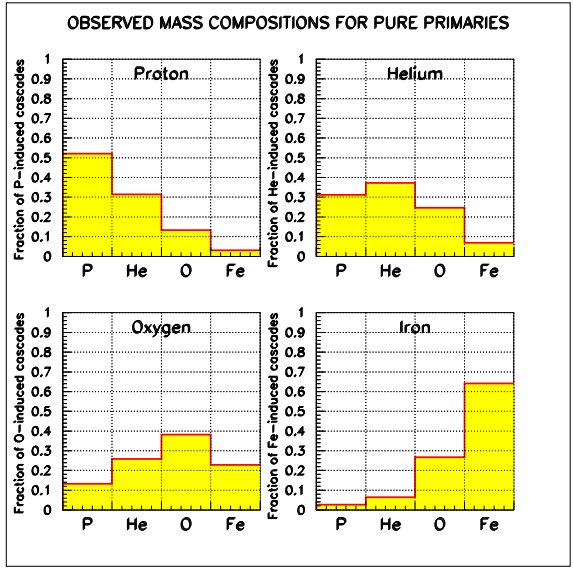}
\caption{\footnotesize Application of the MTA method to the N$_{max}$ -
X$_{max}$ scatter plot - mean probability $P_{ij}$ for the cascades induced by  
{\em i} nucleus: P, He, O and Fe (specified in headers) being identified as induced 
by {\em j} nucleus, indicated at abscissa  \label{fig:mta4}.}
\end{center}
\end{figure}

In order to generate the $X_{max} - N_{max}$ scatter plot and produce the relevant 
matrix of cells we used a set of 4$\times$1000 simulated showers. Then we used another
subset of 4$\times$300 showers to determine the probabilities $P_{ij}$. For each 
individual shower $i$ in a given subset the partial probabilities p$_P^i$ , p$_{He}^i$,
 p$_O^i$ and p$_{Fe}^i$ have been read from the relevant cell {\em i}. The sum of
such probabilities over the entire subset of 300 showers permits
to estimate the probability p$_A$ for a shower of a given nature $A$ to be
identified as a shower generated by the $P$, $He$, $O$ or $Fe$
primary particle. This probability is shown in Figure \ref{fig:mta4}. 
One can see that the method works quite well
being able to attribute the highest probability to the correct nuclei.

\section{\bf Determination of the primary mass composition}

The obtained mean probabilities $P_{ij}$ for
cascades induced by type $i$ nuclei to be identified as those
induced by $j$ nuclei for the pure primary mass composition can be used
for the reconstruction of the mixed primary mass composition as the coefficients in
the system of linear equations:

\begin{eqnarray}
\nonumber n^{\prime}_P & = & \Sigma_{i=1}^4 n_{A_i} \cdot P_{A_i\rightarrow P} \\
n^{\prime}_{He} & = & \Sigma_{i=1}^4 n_{A_i} \cdot P_{A_i\rightarrow He} \\
\nonumber n^{\prime}_O & = & \Sigma_{i=1}^4 n_{A_i} \cdot P_{A_i\rightarrow O} \\
\nonumber n^{\prime}_{Fe} & = & \Sigma_{i=1}^4 n_{A_i} \cdot P_{A_i\rightarrow Fe} 
\end{eqnarray}

where $n_{A_i}$ ($n_P$, $n_{He}$, $n_O$ and $n_{Fe}$) are the true numbers, 
which determine the primary mass composition in the sample of 
$N=n_P+n_{He}+n_O+n_{Fe}$ cascades, which are observed as $n^\prime_P$, 
$n^\prime_{He}$, $n^\prime_O$ and $n^\prime_{Fe}$, due to a misclassification. 

In order to invert the problem and to reconstruct the abundances
$n_{A_i}$ in the primary mass composition from the observed
abundances $n^\prime_{A_i}$ and the known probabilities $P_{ij}$, 
we can apply any method capable to solve the inverse
problem taking into account possible errors of the observed
distribution and the constraint:
\begin{equation}
\Sigma_{i=1}^4 n_{A_i} = \Sigma_{i=1}^4 n_{A_i}^\prime = N
\end{equation}
The observed abundances
were simulated using a set of 4$\times$600 cascades different from those 
used for the determination of $P_{ij}$. At the moment the result of the solution
for the uniform primary mass composition with all abundances 
$p_A= \frac{n_A}{N}$ = 0.25 reconstructed by MTA method is
$p_P = 0.25$, $p_{He} = 0.24$, $p_{O} = 0.23$ 
and  $p_{Fe} = 0.28$.

As an example of the non-uniform mass composition we have taken the case with
$p_P = p_{Fe}$ = 0.1, $p_{He} = p_{O}$ = 0.4, which resembles the
mass composition of cosmic rays at the knee energy of $\sim$3 PeV \cite{Kampe}. 
The total number of 
cascades used for the simulation of non-uniform mass composition is 1500 (150$P$ + 
600$He$ + 600$O$ + 150$Fe$). The result of the 
solution for this case is $p_P = 0.09$, $p_{He} = 0.40$, $p_{O} = 0.35$ 
and  $p_{Fe} = 0.16$.

It is seen that absolute values of the abundance are reconstructed  
satisfactorily.
For the time being the values of 
the probabilities $P_{ij}$ are assumed to be known precisely. A more accurate 
solution of the system, taking propely into account the errors of the 
probability, is under study.

\section{Application for experimental cascades}

It has to be stressed that all the results
outlined above are biased by the fact that we are dealing with
simulated 'noiseless' data and more realistic testing has to be
performed using data which take into account the varying primary
energy, the inclination angle, the instrumental signature (noise)
and the various sources of errors (uneven and incomplete sampling,
etc.). Application of MTA is straightforward - the relevant simulations 
should be made with a maximum possible statistics. 

\section{Conclusions}

We proposed and tested the Multiparametric Topological Analysis method for the 
determination of the {\em mean} primary cosmic ray mass composition on the basis of 
measurements of the longitudinal development of atmospheric cascades. The method 
employs more information about the longitudinal development of atmospheric
cascades than just the depth of the cascade maximum. Definitely the method needs further 
developments before being used for processing real experimental cascades,
but the first results are very encouraging. The next efforts should include the 
increase of simulation statistics, the study of the effect of the shower energy 
spectrum and angular distribution, the accurate estimate of the reconstruction errors.
The great advantage of the proposed MTA method is that it is extremely easy to use 
and generalize for a larger number of observables.

\vspace{5mm}

\noindent{\bf Acknowledgements}
\medskip

\noindent Authors thank M.Risse for simulations of cascades used 
in this work and K.-H.Kampert and L.Perrone for useful discussions.

\end{document}